\documentclass[aps,prl,twocolumn,superscriptaddress,groupedaddress]{revtex4}

\usepackage{graphicx}  
\usepackage{dcolumn}   
\usepackage{bm}        
\usepackage{amssymb}   

\begin{document}

\title{Decoherence Effects on Superpositions of Chiral States in a Chiral Molecule}

\author{M. Bahrami}
\email{mbahrami@ictp.it}
\affiliation{Department of Chemistry, K. N. Toosi University of Technology, 1587-4416 Tehran, Iran}
\affiliation{Condense Matter and Statistical Physics, The Abdus Salam ICTP, Strada Costiera 11, 34151 Trieste, Italy}

\author{A. Shafiee}
\email{shafiee@sharif.edu} \affiliation{Department of Chemistry, Sharif University of
Technology, 11365-9516 Tehran, Iran}

\author{A. Bassi}
\email{bassi@ts.infn.it} \affiliation{Department of Physics, University of
Trieste, Strada Costiera 11, 34151 Trieste, Italy} \affiliation{Istituto
Nazionale di Fisica Nucleare, Trieste Section, Via Valerio 2, 34127 Trieste,
Italy}

\date{\today}
\begin{abstract}
The superposition of chiral states of chiral molecules, as delocalized quantum states of a many-particle system, can be used for the experimental investigations of decoherence theory. In this regard, a great challenge is the precise quantification of the robustness of these superpositions against environmental effects. The methods so far proposed need the detailed specification of the internal states of the molecule, usually requiring heavy numerical calculations. Here, by using the linearized quantum Boltzmann equation and by borrowing ideas employed for analyzing other quantum systems, we present a general and simple approach, of large applicability, which can be used to compute the dominant contribution to the decoherence rate for the superpositions of chiral states of chiral molecules, due to environmental scattering. 
\end{abstract}

\maketitle

\section{Introduction}
Recently there has been great interest in experiments which create the delocalization (superposition of different positions in space) of many-particle complex systems e.g., nano-scale magnets~\cite{magnets}, superconducting rings~\cite{squid}, ensemble  of photons~\cite{photons} or atoms~\cite{atoms}, macro-molecules~\cite{macro} and optomechanical systems~\cite{opto}, with the ultimate aim of testing the validity of quantum linearity~\cite{isart,leg,wein,adler,DP}, and also quantitatively studying decoherence theory~\cite{macro,opto,isart}.
The superposition of chiral states of a chiral molecule is also a very example of spatially-delocalized internal quantum state of many-particle complex systems~\cite{Cin}, thus they can also be used for the aforementioned purposes. The superpositions of chiral states are also very important for other purposes. For example, there are proposals for state-dependent teleportation~\cite{tel}, or the observation of molecular parity violation~\cite{quack}. Meanwhile, the stability of such superpositions and its connection with the optical activity have been the subject of a great debate, usually referred to as Hund paradox~\cite{hund1,hund2,hund3,meanfield1,meanfield2,meanfield3,bahrami}.

Several schemes have been proposed for the preparation of superpositions of chiral states~\cite{superchi1,quack}, e.g., by using ultra-short pulse laser methods~\cite{Cin}.
These superpositions are internal delocalized states which are very sensitive toward localization effects induced by environmental perturbations (e.g., thermal radiation or intermolecular interactions). The quantitative knowledge of these localization effects are of great importance for the aforementioned purposes. Many different approaches have been developed to quantify the environmental effects, the most well-known ones are the mean-field theory~\cite{meanfield1,meanfield2,meanfield3} and the decoherence theory~\cite{hund2,hund3,bahrami}. Very recently, some novel approaches combining the mean-field and the decoherence have been proposed to study the stability of chiral states~\cite{mf_dec}.
Using the mean-field theory, Jona-Lasinio and co-workers developed a very simple method that quantitatively explains, without free parameters, the intermolecular interaction effect on the superposition of chiral states when the pyramidal molecules of the same type are interacting via dipole-dipole interactions~\cite{meanfield2}.
Current proposed methods based on the decoherence theory need the detailed calculations of internal-rotational states of chiral molecule~\cite{hund2,hund3}. These methods usually need heavy machinery numerical calculations for each specific molecule in a particular environment (e.g., see Trost and Hornberger in~\cite{hund3} where the \emph{first} numerical calculation has been performed).

In this paper, by using the well-known linearized quantum Boltzmann master-equation~\cite{JZ,VH,GF,Dios,Adl}, and by borrowing ideas applied to other quantum mechanical systems~\cite{isart,vacc,GF}, we present a very simple and general approach for computing the dominant contribution to the decoherence effect of environmental scattering on the superposition of chiral states in chiral molecules. This approach is of great generality and easy applicability, and does not require knowledge about the detailed structure of the internal quantum states of molecules. As we will show, it will be easily possible to compute the decoherence effects due to the intermolecular interactions and to the blackbody radiation, which are quite common in nature, by determining the most relevant contribution to the corresponding decoherence effects.
We also show that, with a very good approximation, our approach can reproduce the theoretical values obtained by the methods proposed so far in the literature, e.g., the results obtained by Trost and Honberger in~\cite{hund3}. After a brief introduction to Hund's paradox, we also discuss the role of the decoherence due to Cosmic Background Radiation (CBR), sunlight and air molecules on the stability of chiral states of chiral molecules.

\section{Superposition of chiral states as the spatial quantum coherence of a Brownian particle}
Non-planar molecules (including chiral molecules) have at least two equilibrium configurations that can be transformed into each other through inversion around the center-of-mass of the molecule~\cite{Town,Her}. The relevant dynamics can be effectively described by a particle of mass $M$ moving in a double-well potential $V(q)$ with two minima at $q=\pm q_{0}/2$, where $q$ is a generalized inversion coordinate~\cite{Town,Her,Leg,Wei}. The minima, associated with the two localized (say, chiral) states, are separated by a barrier $V_{0}$, as shown in Fig.~1. The biased energy, $\omega_z=V(q_0/2)-V(-q_0/2)$, is the measure of molecular parity violating interactions~\cite{quack,pv_exp}. If we denote the small-amplitude vibration in either well by $\omega_{0}$, then in the limit $ V_{0} \gg \hbar\omega_{0} \gg k_{B}T, $ (where $T$ is the temperature of the bath and $k_{B}$ is Boltzmann constant), the state of the system is effectively confined in the two-dimensional Hilbert space spanned by the localized ground-states of each well, denoted by $|\text{L}\rangle$ and $|\text{R}\rangle$~\cite{Leg,Wei}. For $\omega_z=0$, we denote by $|+\rangle$ and $|-\rangle$ the ground and first excited states ($|\pm\rangle=(|\text{L}\rangle\pm|\text{R}\rangle)/\sqrt{2}$), separated by the energy $E_{-}-E_{+}=\hbar\omega_{x}$, where $\omega_{x}$ is usually called as inversion frequency.

\begin{figure}
  \includegraphics[scale=.8]{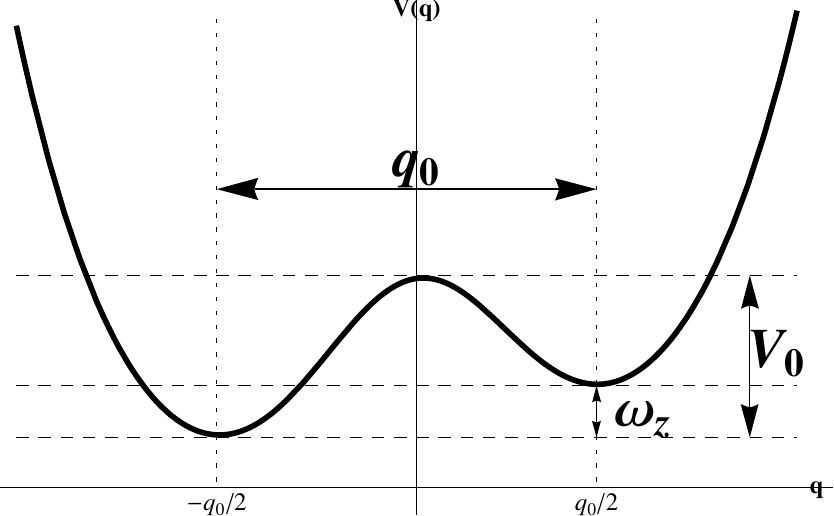}\\
  \caption{An asymmetric double-well potential. $\omega_z$ is the measure of molecular parity violating interactions. The numerically computed value of $\omega_z$ is of order a few Hertz, at best (e.g. $\omega_z = 7.1-9.2\,$Hz for SeOClI)~\cite{quack,pv_exp}. $V_0$ is of order $10^{10}-10^{15}\,$Hz for chiral molecules with stable optical activity~\cite{quack,pv_exp,Town}. The chiral states are localized at each minima, and with a very good approximation are given by the ground states of each minima, if the minima are considered separately.}
  \label{fig:fig1}
\end{figure}

The superposition of chiral states in a chiral molecule is a superposition \emph{in space} of an atom or groups of atoms, between the two minima of the double-well potential. Accordingly, the relevant system can be effectively described as a quantum Brownian particle (or ensemble of particles) of mass $M$ in a superposition of distinct spatial positions, separated by the distance $x=q_0$. The effect of environmental scattering is then manifested as the destruction of this spatial superposition. In~\cite{B2}, we showed how this description works effectively to describe and reproduce the experimental data of the internal tunneling in non-planar molecules. In this article, we generalize this approach.

Following the idea outlined in the previous paragraph, the dynamics of the chiral molecule interacting with a background gas can be described by the linearized quantum Boltzmann master-equation~\cite{JZ,VH,GF,Dios,Adl}.
The main characterization of this master-equation is that the quantum correlation between two different positions ($\langle\textbf{x}|\hat{\rho}(t)|\textbf{x}^\prime\rangle$) decays exponentially with the rate $F(\textbf{x}-\textbf{x}^\prime)$, as follows:
\begin{equation}
\langle\textbf{x}|\hat{\rho}(t)|\textbf{x}^\prime\rangle \propto
e^{-F(\textbf{x}-\textbf{x}^\prime)\,t} \,\langle\textbf{x}|\hat{\rho}(0)|\textbf{x}^\prime\rangle
\end{equation}
with $\hat{\rho}$ the corresponding density matrix of system. The complete formula of $F(\textbf{x}-\textbf{x}^\prime)$ can be very complicated~\cite{JZ,Adl,VH,Dios,GF}. A first simplifications comes from considering isotropic scattering, in which case one has: $F(\textbf{x}-\textbf{x}^\prime)=F(x)$ with $|\textbf{x}-\textbf{x}^\prime|=x$. In general, $F(x)$ has two asymptotic behaviors~\cite{Adl,vacc,GF,isart}:
(1) The saturation behavior at large scales, and (2) The quadratic behavior at small scales, which are expressed mathematically as follows:
\begin{equation}\label{eq:asymp}
    F(x\rightarrow\infty)=\gamma, \qquad F(x\rightarrow0)=\Lambda\, x^2
\end{equation}
Having these asymptotic behaviors in mind, a very resonable ansatz for $F(x)$ can be given by (see Refs.~\cite{isart,vacc,GF} for more detail):
\begin{equation}
\label{eq:F1}
F(x) \equiv \gamma\,(1-\exp[-x^2/4r^2])
\end{equation}
with $\gamma$ the localization strength, and $r$ the localization
distance. As we see, for $x\gg2r$, one gets $F(x)=\gamma$, and for $x\ll2r$ one has $F(x)=(\gamma/4r^2)x^2$, and thus $\Lambda=\gamma/4r^2$.
Therefore, $F(x)$, as given by Eq.~(\ref{eq:F1}), is determined by only two parameters: $\gamma$ and $r$.
Henceforth, we will apply this simple and general ansatz to describe the localization rate of chiral states in chiral molecules. In the next section, we will provide an expression for $\gamma$ and $r$ derived from the underlying microscopic dynamics. We will investigate the decoherence effect of the blackbody radiation and the intermolecular interactions in more detail.

\section{A master formula for localizing effect by environmental scattering}
As explained before, we have considered the superposition of chiral states in chiral molecules as a spatial superposition of a Brownian particle in two different positions in space, with superposition distance $x=q_0$. Accordingly, one can apply the linearized quantum Boltzmann master-equation to compute the localization effect of the environmental scattering. The linearized quantum Boltzmann master-equation has been elaborated in the literature (e.g., see Refs.\cite{JZ,Adl,VH,Dios}). The corresponding formulations were initiated by Joos and Zeh~\cite{JZ}, and the general and corrected formula was given by Di\'{o}si~\cite{Dios} (see Refs.\cite{Adl,VH} for more detail). We work in dilute gas limit where three-body collisions or correlated scatterings are negligible. We also assume a bath of uniform density. We take the scattering to be recoil-free. For isotropic scattering, we may apply the following mathematical form of the localization rate, which is more suitable for our purposes~\cite{Adl,VH}:
\begin{eqnarray}\label{eq:F}
    F(x)& = & \int_0^\infty dp \, \varrho(p) \, v(p) \, \{ \sigma_t(p)-
    \\ \nonumber
    &&\qquad\qquad\qquad
    2\pi \int_{-1}^{+1} d(\cos \theta)
     \frac{\sin \Theta}{\Theta}\left|f\left(p,\theta\right)\right|^{2}\}
\end{eqnarray}
with $v$ [$p$] the bath particle velocity [momentum],
$\varrho(p)$ the normalized number density of bath particles' momenta,
$f\left(p,\theta\right)$ the scattering amplitude,
$\theta$ the scattering angle,
$\Theta=(2xp/\hbar)\sin\left(\theta/2\right)$, and
$\sigma_t(p)=2\pi\int^{+1}_{-1} d\left(\cos\theta\right)\left|f\left(p,\theta\right)\right|^{2}$
the total cross-section.
Regarding asymptotic behaviors of Eq.(\ref{eq:F}), one finds:
\begin{eqnarray}
\label{eq:gamma}
\gamma& = & \lim_{x\to\infty}F(x)=\int_0^\infty dp \, \varrho(p) \, v(p) \, \sigma_t(p)
\\
\label{eq:r}
r & = &\frac{\sqrt{\gamma}}{2}\lim_{x\to 0}\sqrt{\frac{x^2}{F(x)}}
\\\nonumber
&=&\frac{\hbar\sqrt{3\gamma}}{2}
\left(\int_0^\infty dp\, \varrho(p)\, v(p)\, p^2 \sigma_s(p)
\right)^{-\frac{1}{2}}
\end{eqnarray}
with $\sigma_s(p)=4\pi\int^{+1}_{-1} d\left(\cos\theta\right) \sin^2(\theta/2)\left|f\left(p,\theta\right)\right|^{2}$.

Eqs.~(\ref{eq:gamma}) and~(\ref{eq:r}) allows us to compute the localizing effects induced by many different scattering processes, e.g., scattering by a background gas (i.e., the decoherence effects of intermolecular interactions) or by the blackbody radiation. In the following, we will find the analytical expressions for these special environments, which are quite common in many physical situations.

\subsection{Scattering by thermal radiation}
Using the Rayleigh scattering~\cite{jack} (or Mie scattering, in general~\cite{scat}) and then combining with Eq.(\ref{eq:gamma}) and~(\ref{eq:r}), we can compute the localizing effects of elastic scattering of a blackbody radiation. We consider the Brownian particle as a small dielectric sphere of radius $a$, with the uniform isotropic dielectric constant $\epsilon$. Accordingly, when $\zeta_{\text{th}}\gg a$, with $\zeta_{\text{\tiny th}}$ the thermal wavelength of light, one has:
$\sigma_s(p)=\sigma_t(p)=\frac{8\pi}{3}(p/\hbar)^4a^6
|(\varepsilon-1)/(\varepsilon+2)|^2$~\cite{jack}.
Introducing this into Eqs.(\ref{eq:gamma}) and~(\ref{eq:r}) and using the Planck distribution $\varrho(p)=p^2/\left(\pi^2\hbar^3(\exp[cp/k_BT]-1)\right)$, one gets:
\begin{eqnarray}
\gamma_{\text{\tiny TR}}&=&\frac{5760\,\xi(7)}{3\pi}a^6\,c\,(\frac{k_BT}{c\hbar})^7
\left|\frac{\varepsilon-1}{\varepsilon+2}\right|^2
,
\\
r_{\text{\tiny TR}}&=&\sqrt{\frac{3\hbar\,\xi(7)}{224\,\xi(9)}}
\frac{c\hbar}{k_BT}
\end{eqnarray}
where $\xi(x)$ is Riemann zeta function.

As an example, we can compute the decoherence effects induced by Cosmic Background Radiation (CBR), which is equivalent to a blackbody radiation of $T=2.735\,$K. We consider our system of radius $a=a_{\text{\tiny Bohr}}$ (with $a_{\text{\tiny Bohr}}$ the Bohr radius) and $|(\varepsilon-1)/(\varepsilon+2)|\approx1$. Accordingly, one gets:
$\gamma_{\text{\tiny TR}}\simeq10^{-29}\,$Hz, $r_{\text{\tiny TR}}\simeq10^{-4}\,$m, and therefore, according to Eq.~(\ref{eq:F1}), $F_{\text{\tiny TR}}(q_0)\simeq 10^{-42}\,$Hz for $q_0=a_{\text{\tiny Bohr}}$. Likewise, the sunlight on earth can be envisaged as a blackbody radiation of the temperature around $T=5523\,$K. Accordingly, for sunlight, one obtains:
$\gamma_{\text{\tiny TR}}=10^{-5.7}\,$Hz, $r_{\text{\tiny TR}}=10^{-7.3}\,$m, and Eq.~(\ref{eq:F1}) gives: $F_{\text{\tiny TR}}(q_0)\simeq 10^{-12.2}\,$Hz for $a=q_0=a_{\text{\tiny Bohr}}$.
Using Wien's law, one gets $\zeta_{\text{\tiny th}}\approx10^{-4}\,$m and $\zeta_{\text{\tiny th}}\approx10^{-7}\,$m for the CBR and sunlight radiations, respectively, which confirms the validity for Rayleigh scattering (i.e., $\zeta_{\text{\tiny th}}\gg a$), since $a$ has the order of few Angstroms.

\subsection{Scattering by a background gas}
In the case of a gaseous environment, one can relate the scattering amplitude, $f(p,\theta)$, to the scattering potential by using the phase-shift methods~\cite{Joach}. If we consider the scattering interaction as given by a spherically symmetric potential $V(r')$ (with $r'$ the distance between center-of-masses of colliding molecules), then $f\left(p,\theta\right)=
(\hbar/p)\sum_{\ell=0}^{\infty}\left(2\ell+1\right)e^{i\delta_{\ell}(p)}
\sin{\delta_{\ell}(p)}\:P_{\ell}\left(\cos{\theta}\right)$,
with $P_{\ell}\left(\cos{\theta}\right)$ the Legendre polynomials, and $\delta_{\ell}$ the phase shifts depending on the scattering potential. Consequently, one gets:
\begin{eqnarray}
\label{eq:sigma}
\sigma_t(p)=\frac{4\pi \hbar^2}{p^2}S_1(p),
\qquad\quad
\sigma_s(p)=\frac{4\pi \hbar^2}{p^2}S_2(p),
\end{eqnarray}
where
\begin{eqnarray}
S_1(p)&=&\sum_{l=0}^{\infty}(2l+1)\sin^2\delta_l(p)
\\
S_2(p)&=&\sum_{l=0}^{\infty}
(2l+1)\sin^2\delta_l(p)
\\\nonumber
&+&\sum_{l=0}^{\infty}2(l+1)\cos(\delta_{l+1}(p)-\delta_{l}(p))
\sin\delta_{l+1}(p) \sin\delta_{l}(p)
\end{eqnarray}

Accordingly, if we know the interaction potential, we can compute the scattering amplitude and thus we can compute the localization rate, too. Considering the Brownian particle and the background gas as two spheres, then we can assume that they behave like the atoms of rare gases interacting with the potential of the type: $V(r')=-C_s/r'^s$~\cite{mait}. We can employ the Jeffereys-Born approximation to compute the phase-shifts in Eq.~(\ref{eq:sigma})~\cite{Joach}:
\begin{equation}\label{eq:BJ}
    \delta_l(p)\approx\frac{\pi m C_s}{\hbar p}\left(\frac{p}{2\hbar}\right)^{s-1}
    \frac{\Gamma(s-1)\,\Gamma(l-s/2+3/2)}{[\Gamma(s/2)]^2\,\Gamma(l+s/2+1/2)}
\end{equation}
with $\Gamma(x)$ the gamma function.

If we assume that $\sigma_t(p)$ and $\sigma_s(p)$ do not change appreciably for all important bath particles' momenta $\bar{p}$, then we can approximate Eqs.~(\ref{eq:gamma}) and~(\ref{eq:r}) as follows:
\begin{eqnarray}  \label{eq:BG}
  \gamma_{\text{\tiny BG}} \simeq n \frac{4\pi \hbar^2}{m\bar{p}}S_1(\bar{p}),
  \quad
  r_{\text{\tiny BG}} \simeq \frac{\hbar}{\bar{p}}\sqrt{\frac{3S_1(\bar{p})}{S_2(\bar{p})}},
\end{eqnarray}
where $n$ is the density of the background gas. One can simply replace $\bar{p}$
with the most probable momentum of a Maxwell-Boltzmann gas: $\bar{p}=\sqrt{2mk_BT}$; also $n$ can be replaced by the density of an ideal gas: $n=P/k_BT$.

To give an estimate of above formula, let us consider a chiral molecule immersed in the air molecules ($m=28\,$amu) as background gas at room temperature. We assume the chiral molecule interacts with the bath particle via London dispersion interaction ($s=6$).
Since the London dispersion coefficient is in the range $C_6\sim10^{-79}-10^{-76}\,$J$\cdot$m$^6$~\cite{mait}, then one finds
$r_{\text{\tiny BG}}\simeq 10^{-11.2}\,$m. For $q_0=a_{\text{\tiny Bohr}}$, one gets from Eq.~(\ref{eq:F1}):
$F_{\text{\tiny BG}}(q_0)\simeq 10^{9.8}-10^{11.1}\,$Hz in the standard pressure, and $F_{\text{\tiny BG}}(q_0)\simeq 10^{-3.5}-10^{-2.3}\,$Hz in the the laboratory vacuum ($n\sim10^{12}\,$particle/m$^3$).

Likewise, for Helium as the background gas at room temperature and $q_0$ in the order of few Angstroms, one finds: $F_{\text{\tiny BG}}(q_0)\simeq 10^5(P/\text{Pa})-10^6(P/\text{Pa})\,$Hz, with $P$ the pressure of Helium.
In~\cite{hund3}, Trost and Honberger have computed the decoherence rate (denoted by $\lambda$) on the internal tunneling of D$_2$S$_2$ due to the collisions with Helium as the background gas. After quite involved numerical calculations, they obtained $\lambda\approx10^{2.4}\,$Hz at $T=300\,$K and $P=1.6\times10^{-3}\,$Pa. Introducing these values for $T$ and $P$ into Eq.~(\ref{eq:BG}) and combing the result with Eq.~(\ref{eq:F1}), one gets: $\lambda\approx F_{\text{\tiny BG}}(q_0)\approx10^{2.7}\,$Hz (where we have used $q_0=1.9$\AA$\;$ and $C_6\simeq10^{-78}\,$J$\cdot$m$^6$~\cite{hund3,B2}), which is almost identical to the result of~\cite{hund3}. This shows that the approach here proposed can provide a good estimate of the order of magnitude of the decoherence rate due to environmental scattering.

A further comment is at order. The values we have obtained for $F(q_0)$ for air molecules at room temperature and for the sunlight are much greater than the inversion frequency of a typical chiral molecule with the stable optical activity ($\omega_x\sim10^{-13}-10^{-80}\,$Hz~\cite{quack,berger}).
Accordingly, since the decoherence rate is much greater than the inversion frequency (i.e., $F(q_0)\gg\omega_x$), then the superposition of chiral states of optically-stable chiral molecules (even if they are produced in a particular natural process) are not stable in the atmospheric environment of Earth, and their life-time is of the order of $\sim10^{-9.8}-10^{-11.1}\,$s.
Thus, the expected natural habitant states of optically-stable chiral molecules in the atmospheric environment of Earth are chiral states, as confirmed by observations.

In addition, our calculations also show that the \emph{inevitable} localization effect of CBR is also greater than $\omega_x$ of some chiral molecules, in particular those with very stable optical activity, like some complex and biologically-important chiral molecules. In this case, since $F(q_0)\gg\omega_x$, then the superposition of chiral states are unstable, and the chiral states become stabilized. Thus, for optically-stable chiral molecules whose inversion frequency is much smaller than the decoherence effect of CBR
(i.e., $F_{\text{\tiny CBR}}\sim10^{-42}\gg\omega_x$), the superposition of chiral states are not stable, the environment stabilizes the chiral states. A good approximation for the inversion frequency is given by $\omega_x\approx A\;\omega_0\sqrt{\frac{V_0}{\hbar\omega_0}}\,\text{exp}[-B\frac{V_0}
{\hbar\omega_0}]$ with $A,B$ of the order of unity, depending on the explicit mathematical form of the double-well potential~\cite{Wei}. Since the small-amplitude frequency, $\omega_0$, which is the typical molecular vibrations, is in the range $\omega_0\sim100-1000\,$cm$^{-1}$, then for $F_{\text{\tiny CBR}}\sim10^{-42}\gg\omega_x$ one finds $V_0\geq10^{13}\,$Hz. In other word, for the molecules with this range of inversion barrier, the CBR may stabilize the chiral states. This may be considered as a partial explanation for Hund's paradox, which arises because optically-stable chiral molecules are in definite chiral states, either $|\text{L}\rangle$ or $|\text{R}\rangle$, although these states are not eigenstates of the Hamiltonian when $\omega_x\gg\omega_z$~\cite{hund1,hund2,hund3,bahrami}.
However, for typical carbon-based biomolecules such as amino acids, the sophisticated calculations show that $\omega_z$ has typically an order of magnitude larger than $10^{-4}-10^{-3}\,$Hz~\cite{quack,pv_exp}, meaning that the parity violating interactions (i.e. $\omega_z$) are the dominant term that stabilize the chiral states.
In other word, the decoherence effect of CBR can explain the Hund paradox for those molecules only if $F_{\text{\tiny CBR}}\gg\omega_z$.

\section{Conclusion}
The superposition of chiral states of a chiral molecule are very important for the practical and also fundamental research in physics and chemistry.
A great challenge is the quantification of robustness of these superpositions against environmental effects.
In this paper, by using the quantum Brownian motion approach, we provided a formula for computing the dominant contribution to the decoherence rate. This formula is of general applicability and simplicity compared with other methods. In this approach, there is no need to know the detailed internal structure of the molecules (e.g., their angular momentum states), and the mere knowledge of the distance of the minima of double-well ($q_0$), the scattering potential and the thermodynamical variables of bath (e.g., temperature and pressure) are sufficient in order to compute the dominant contribution to the localization rate induced by the environmental scattering.

We applied the formalism, and computed the decoherence rate for two quite common types of environment: thermal radiation and background gas. We have showed how our results are almost identical to the result of the other methods proposed so far. We have commented how environmental decoherence can explain why the superpositions of chiral states of chiral molecules are not stable in the atmospheric environment of Earth.

\textbf{Acknowledgments:}
M. Bahrami acknowledges hospitality from The Abdus Salam ICTP, where this work was carried out, and also the partial financial support from Iran National Elite Foundation (BMN), K. N. Toosi University of Technology, and Iran National Science Foundation (INSF).
A. Shafiee acknowledges partial financial support from Sharif University of Technology, and Iran National Science Foundation (INSF).
A. Bassi acknowledges partial financial support from MIUR (PRIN 2008), INFN, and COST (MP1006).


\begin{thebibliography}{99}

\bibitem{magnets}
J. R. Friedman {\it et al.} Phys. Rev. Lett. \textbf{76}, 3830-3 (1996).
del Barco {\it et al.} Europhys. Lett. \textbf{47}, 722-8 (1999).
Wernsdorfer {\it et al.} Phys. Rev. Lett. \textbf{79}, 4014-7 (1997).

\bibitem{squid}
Y. Nakamura {\it et al.} Nature \textbf{398}, 786-788 (29 April 1999).
C. H. van der Wal {\it et al.}, Science \textbf{290}, 773 (2000).
J. R. Friedman {\it et al.}, Nature (London) \textbf{406}, 43 (2000).

\bibitem{photons}
S. Delglise {\it et al.} Nature \textbf{455}, 510-14 (2008).

\bibitem{atoms}
K. Hammerer {\it et al.} Rev. Mod. Phys. \textbf{82}, 1041-1093 (2010).

\bibitem{macro}
M. Arndt {\it et al.}, Nature {\bf 401}, 680 (1999).
S. Gerlich {\it et al.}, Nature Phys. {\bf 3}, 711 (2007).
S. Gerlich {\it et al.}, Nature Comm. {\bf 2}, 263 (2011).
O. Romero-Isart {\it et al.}, New J. Phys. \textbf{12}, 033015 (2010).
O. Romero-Isart {\it et al.}, Phys. Rev. Lett. \textbf{107}, 020405 (2011). arXiv:1103.408

\bibitem{opto}
P. F. Cohadon, A. Heidmann, and M. Pinard, Phys. Rev. Lett. {\bf 83}, 3174 (1999).
Marshall \emph{et. al.} Phys. Rev. Lett. \textbf{91}, 130401 (2003).
O. Arcizet {\it et al.}, Nature {\bf 444}, 71 (2006).
Kippenberg and Vahala Science \textbf{321}, 1172�6 (2008).
Marquardt and Girvin Physics \textbf{2} 40, (2009).
Favero and Karrai Nature Photon. \textbf{3}, 201�5 (2009).

\bibitem{isart}
Oriol Romero-Isart, Phys. Rev. A \textbf{84}, 052121 (2011).
Nimmrichter \emph{et al.} Phys. Rev. A \textbf{83}, 04362 (2011).

\bibitem{leg}
J. A. Leggett, Suppl. Prog. Theor. Phys. \textbf{69}, 80 (1980); Phys.: Condens. Matter \textbf{14}, R415 (2002).

\bibitem{wein}
S. Weinberg, Phys. Rev. Lett. \textbf{62}, 485 (1989); Ann. Phys. \textbf{194}, 336 (1989);
Nuclear Physics B 6, 67 (1989); arXiv:1109.6462 (2011).

\bibitem{adler}
S. L. Adler, arXiv: quant-ph/0004077 (2000); Quantum Theory as an Emergent Phenomenon, (Cambridge University Press, Cambridge, UK, 2004).
S. L. Adler, and A. Bassi, Science \textbf{325}, 275 (2009);

\bibitem{DP}
R. Penrose, Gen. Relativ. Gravit. \textbf{28}, 581 (1996).
L. Diosi, J. Phys. A: Math. Theor. \textbf{40}, 2989 (2007); Phys. Lett. A \textbf{105}, 199 (1984).

\bibitem{Cin}
J. A. Cina and R. A. Harris, Science \textbf{267}, 832 (1995).

\bibitem{tel}
C. S. Maierle \emph{et al.}, Phys. Rev. Lett. \textbf{81}, 5928 (1998).

\bibitem{quack}
M. Quack, Chem. Phys. Lett. \textbf{132} (1986) 147.
M. Quack, Angew. Chem. Int. Ed. Engl. \textbf{28} (1989) 571.
A. Bakasov, T.-K. Ha, M. Quack, J. Chem. Phys. 109 (1998) 7263.
M. Quack, Angew. Chem. Intl. Ed. (Engl.) \textbf{41} (2002) 4618.
M. Quack, J. Stohner, Chimia \textbf{59} (2005) 530.
M. Quack, J. Stohner, M. Willeke, Annu. Rev. Phys. Chem. \textbf{59} (2008) 741.

\bibitem{meanfield1}
A. Vardi: J. Chem. Phys. \textbf{112}, 8743 (2000).
\bibitem{meanfield2}
G. Jona-Lasinio, C. Presilla, C. Toninelli: Phys. Rev. Lett. \textbf{88}, 123001 (2002).
\bibitem{meanfield3}
V. Grecchi, A. Sacchetti: J. Phys. A: Math. Gen. \textbf{37}, 3527 (2004).

\bibitem{hund1}
W. Hund: Z. Phys. \textbf{40}, 742 (1927); Z. Phys. \textbf{43}, 805 (1927).
E. Merzbacher: \textit{Physics Today} 44 (Aug. 2002).

\bibitem{hund2}
R. A. Harris, L. Stodolsky, Phys. Lett. B \textbf{78}, 313 (1978); J. Chem. Phys. \textbf{74}, 2145 (1981); Phys. Lett. B \textbf{116}, 464 (1982); J. Chem. Phys. \textbf{78}, 7330 (1983).
R. Silbey, R. A. Harris, J. Chem. Phys. \textbf{80}, 2615 (1984); J. Phys. Chem. \textbf{93}, 7062 (1989).
M. Simonius: Phys. Rev. \textbf{40}, 980 (1978).
P. Pfeifer: Phys. Rev. A. \textbf{26},701 (1982).
Y. A. Berlin \textit{et. al.}: Z. Phys. D 37, 333-339 (1996).
\bibitem{hund3}
J. Trost, K. Hornberger: Phys. Rev.Lett. \textbf{103}, 023202 (2009).

\bibitem{bahrami}
M. Bahrami, A. Shafiee, Comput. Theoret. Chem. \textbf{978} (2011) 84-87.

\bibitem{superchi1}
J. A. Cina and R. A. Harris, J. Chem. Phys. \textbf{100}, 2531 (1994).

\bibitem{mf_dec}
I. Gonzalo and P. Bargueo, Phys. Chem. Chem. Phys. \textbf{13} (2011) 17130-17134
P. Bargueno \emph{et al.} Chem. Phys. Lett. \textbf{516} (2011) 29-34.

\bibitem{JZ}
E. Joos and H. D. Zeh: Z. Phys. B \textbf{59}, 223 (1985).

\bibitem{Dios}
L. Diosi, Europhys. Lett. \textbf{30}, 63 (1995).

\bibitem{Adl}
S. L. Adler, J. Phys. A: Math. Gen. \textbf{39}, 14067�14074 (2006).

\bibitem{VH}
B. Vacchini and K. Hornberger, Phys. Rep. \textbf{478}, 71�120 (2009).

\bibitem{GF}
M. R. Gallis and G. N. Fleming, Phys. Rev. A 42, 38 (1990).

\bibitem{vacc}
B. Vacchini, J. Phys. A: Math. Theor. 40, 2463 (2007)

\bibitem{Town}
C. H. Townes and A. L. Schawlow, \textit{Microwave Spectroscopy}, (McGraw-Hill, New York, 1955)

\bibitem{Her}
G. Herzberg:  \textit{Molecular Spectra and Molecular Structure. Electronic Spectra and Electronic Structure of Polyatomic Molecules} (Krieger: Malabar, FL, Reprint 1991) Vol. III

\bibitem{Leg}
A. J. Leggett \textit{et. al.}: Rev. Mod. Phys. \textbf{59}, 1 (1987)

\bibitem{Wei}
U. Weiss, \emph{Quantum Dissipative Systems}, World Scientific, Singapore, 2008.


\bibitem{pv_exp}
A. J. MacDermott, Enantiomer 5 (2000) 153.
R. Zanasi, P. Lazzeretti, Chem. Phys. Lett. 286 (1998) 240.
F. Faglioni, P. Lazzeretti, Phys. Rev. A \textbf{67} (2003) 0321011-4.
P. Schwerdtfeger, J. Gierlich, T. Bollwein, Angew. Chem. Int. Ed. Engl. \textbf{42} (2003)
1293-1296.
R. Bast, P. Schwerdtfeger, Phys. Rev. Lett. \textbf{91} (2003) 0230011-3.
F. De Montigny \emph{et. al}, Phys. Chem. Chem. Phys. \textbf{12} (2010) 8792.
D. Figgen, T. Saue, P. Schwerdtfeger, J. Chem. Phys. \textbf{132} (2010) 234310.
B. Darqui\'{e} \emph{et al.}, Chirality \textbf{22} (2010) 870.

\bibitem{B2}
M. Bahrami, A. Bassi, Phys. Rev. A. \textbf{84}, 062115 (2011).

\bibitem{jack}
J. D. Jackson, \emph{Classical elecrodynamics}, 3rd Ed. (Wiley, New York 1998) p.460.

\bibitem{scat}
H. C. van de Hulst, \emph{Light Scattering by Small Particles} (Wiley, New York,
1957).

\bibitem{Joach}
C. J. Joachain, \emph{Quantum Collision Theory} (Elsevier Science Publisher B.V., Amsterdam, 1975), Chapter 4.
 M. S. Child, \emph{Molecular Collision Theory} (Academic Press, London, 1984).

\bibitem{mait}
C. G. Miatland \emph{et al.}, \emph{Intermolecular Forces Intermolecular forces: their origin and determination} (Clarendon Press 1981).

\bibitem{berger}
R. Berger \emph{et al.}, Angew. Chem. Int. Ed. \textbf{44}, 3623 (2005)

\end{thebibliography}
\end{document}